\documentclass[a4paper,12pt]{article}
\usepackage{graphicx}
\usepackage{amsmath}
\usepackage{a4}
\usepackage{times}

\input{tcilatex}

\begin{document}

\title{The form factor of the transition $\gamma \gamma ^{\ast }\pi ^{0}$ }
\author{N. F. Nasrallah\thanks{%
email: nsrallh@ul.edu.lb} \\
Institut de Physique Nucl\'{e}aire, Universit\'{e} Paris XI, \\
F-91406 Orsay Cedex, France\\
and Faculty of Science, Lebanese University, \\
Tripoli, Lebanon}
\date{}
\maketitle

\begin{abstract}
A simple method is used to extrapolate the form factor for the process $%
\gamma ^{\ast }\gamma ^{\ast }\rightarrow \pi ^{0}$ when two virtual photons
produce a $\pi ^{0}$ from the region of large spacelike photon virtualities
to the experimentally accessible case when one of the photons is real. The
method is also applied to the study of the axial anomaly and $\omega -\rho
-\pi $ coupling and to the pion charge and charge radius. The results
compare favourably with the experimental measurements.
\end{abstract}

\section{Introduction}

The study of the process $\gamma ^{\ast }\gamma ^{\ast }\rightarrow \pi ^{0}$
and the related form factors when two virtual photons produce a neutral pion
provides a simple, interesting and clean test of QCD. In contrast to the
pion electromagnetic form factor it is not asymptotically supressed by being
multplied by the strong coupling constant $\alpha _{s}$. In the lowest order
in $\alpha _{s}$, perturbative QCD predicts that \cite{1} 
\begin{equation}
F_{\gamma ^{\ast }\gamma ^{\ast }\rightarrow \pi ^{0}}^{LO}(q^{2},Q^{2})=%
\frac{4\pi }{3}\int_{0}^{1}\frac{\phi _{\pi }(x)}{xQ^{2}+\bar{x}q^{2}}dx 
\tag{1.1}  \label{1.1}
\end{equation}
where $\phi _{\pi }(x)$ is the phenomenological pion distribution amplitude
and $x,\;\bar{x}=1-x$ are the fractions of the pion light cone momenta
carried by the quarks in the region where both photon virtualities $%
q^{2}=-q_{1}^{2}$ and $Q^{2}=-q_{2}^{2}$ are large.

The most experimentally favourable situation is when one of the photons is
real, $q^{2}=0$ and this has led to the measurement of the form factor $%
F_{\gamma \gamma ^{\ast }\pi ^{0}}(q^{2}=0,Q^{2})$ by the CELLO \cite{2} and
CLIO \cite{3} groups.

Recently, Radyushkin and Ruskov \cite{4} have undertaken to obtain the
transition form factor $F_{\gamma \gamma ^{\ast }\pi ^{0}}(q^{2}=0,Q^{2})$
independently of the specific form of the pion distribution amplitude. They
\ have rather first evaluated the perturbative and non-perturbative
contributions to the amplitude $F_{\gamma ^{\ast }\gamma ^{\ast }\rightarrow
\pi ^{0}}(q^{2},Q^{2})$ in the deep Euclidean region and have the
extrapolated to the mass-shell of one of the photons in a QCD\ sum rule \cite
{5} approach. It is the purpose of this note to undertake this extrapolation
in a more straightforward and much less model dependant way.

The task of extrapolating QCD\ prodictions from the deep asymptotic region
to low energies is not a trivial one, an effective way is the use of QCD\
sum rules \cite{5} in which use is made of the analyticity properties of an
amplitude to write dispersion integrals the low energy part of which is
saturated by known resonances and where QCD is employed to evaluate the
asymptotic part. It is in general hard to assess the quality of the
approximations made as resonance saturation should not be expected to hold
over too large an interval and deviations from QCD could become important
far away from the Euclidean region.

A simple model independant method of extrapolation is proposed in this note
in which the integration kernel in the dispersion integral is modified by
adding to it polynomials of low order $N$. The polynomials are chosen such
as to annihilate the kernel as well as its derivatives up to order $N$ in
the region where both resonance saturation and QCD are least reliable.

The method is applied to obtain the form factor $F_{\gamma \gamma ^{\ast
}\pi ^{0}}(q^{2}=0,Q^{2})$. It is shown that extrapolation of the QCD
expression of Radyushkin and Ruskov \cite{4} with the values they use for
the damping parameters, which they deduce from stability considerations, as
well as the values commonly used for the QCD non-perturbative vacuum
condensates\footnote{%
The result turn out to be insensitive to the exact values of these.} lead to
values of $F_{\gamma \gamma ^{\ast }\pi ^{0}}(q^{2}=0,Q^{2})$ in good
agreement with the experimental data.

Another application of the method is to the anomaly relation \cite{6} $%
F_{\gamma \gamma \pi ^{0}}=\frac{1}{\sqrt{2}\pi f_{\pi }}$, where $f_{\pi
}=92.4\mathrm{MeV}$ is the pion decay constant, which holds when both
photons are on their mass-shell. In particular we evaluate corrections to
the value of the $\omega -\rho -\pi $ coupling constant given by vector
meson dominance.

The effectiveness and stability of the method is finally demonstrated by
calculating the charge and charge radius of the pion from the asymptotic
behaviour of the pion's electromagnetic form factor obtained some time ago
by Ioffe and Smilga \cite{7} and by Nesterenko and Radyushkin \cite{8}.
Agreement with experiment is good in all cases.

\section{The $\protect\gamma \protect\gamma ^{\ast }\protect\pi ^{0}$ form
factor}

The electroproduction of a neutral pion in the reaction $\gamma ^{\ast
}\rightarrow \gamma \pi ^{0}$ is experimentally accessible \cite{2}, \cite{3}%
. The corresponding form factor is defined when both photons are off their
mass-shell. 
\begin{equation}
\sqrt{2}i\epsilon _{\mu \nu \alpha \beta }q_{1}^{\alpha }q_{2}^{\beta
}F_{\gamma ^{\ast }\gamma ^{\ast }\pi ^{0}}(q_{1}^{2},q_{2}^{2})=4\pi \int
dx\,\exp (iq_{1}x)\langle \pi (p)\mid TV_{\mu }(x)V_{\nu }(0)\mid 0\rangle 
\tag{2.1}  \label{2.1}
\end{equation}
where $V_{\mu }$ is the electromagnetic current of light quarks divided by
the unit charge 
\begin{equation*}
V_{\mu }=(\frac{2}{3}\bar{u}\gamma _{\mu }u-\frac{1}{3}\bar{d}\gamma _{\mu
}d)
\end{equation*}

Radyushkin and Ruskov \cite{4} calculate the QCD expression of the form
factor which is valid when both $t=q_{1}^{2}$ and $Q^{2}=-q_{2}^{2}$ are in
the deep euclidean region 
\begin{equation}
F^{\mathrm{QCD}}(t,Q^{2})=\sqrt{2}\pi f_{\pi }F_{\gamma ^{\ast }\gamma
^{\ast }\pi ^{0}}^{\mathrm{QCD}}(t,Q^{2})=F_{p}(t,Q^{2})+c_{-2}t+c_{0}+\frac{%
c_{2}}{t}+\frac{c_{4}}{t^{2}}+\cdots  \tag{2.2}  \label{2.2}
\end{equation}
with the perturbative part 
\begin{equation}
F_{p}(t,Q^{2})=-2\int_{0}^{R_{0}}ds\,\exp \left( -\frac{s}{M^{2}}\right)
\int_{0}^{1}dx\,\frac{\bar{x}}{x^{2}}\frac{(\bar{x}Q^{2}-xt)^{2}}{\left[ t-%
\frac{\bar{x}}{x}\left( xs+Q^{2}\right) \right] ^{3}}  \tag{2.3}  \label{2.3}
\end{equation}

$M^{2}$ is the Borel damping mass and $R_{0}$ is the duality radius in the
axial-vector channel. Both are determined by stability considerations \cite
{7}, \cite{8}, the values $M^{2}\simeq 1\mathrm{GeV}^{2}$ and $R_{0}=.7%
\mathrm{GeV}^{2}$ are used. The coefficients 
\begin{eqnarray}
c_{-2}(M^{2},Q^{2}) &=&-\frac{64}{243}\pi ^{3}\frac{\alpha _{S}\langle
(qq)^{2}\rangle }{M^{2}Q^{4}}  \notag \\
c_{0} &=&\left[ \frac{\pi ^{2}}{18M^{2}}\langle \frac{\alpha _{S}GG}{\pi }%
\rangle +\frac{32}{27}\frac{\pi ^{3}}{M^{4}}\alpha _{S}\langle
(qq)^{2}\rangle \right]  \notag \\
c_{2} &=&-\left[ \frac{\pi ^{2}}{9}\langle \frac{\alpha _{S}GG}{\pi }\rangle
\left( \frac{1}{2M^{2}}-\frac{1}{Q^{2}}\right) +\frac{64}{27}\pi ^{3}\alpha
_{S}\langle (qq)^{2}\rangle \left( \frac{1}{2M^{4}}-\frac{1}{Q^{4}}\right) %
\right]  \notag \\
c_{4} &=&\frac{64\pi ^{3}}{243}\alpha _{S}\langle (qq)^{2}\rangle \left( 
\frac{Q^{2}}{M^{4}}+\frac{q}{Q^{2}}\right)  \TCItag{2.4}  \label{2.4}
\end{eqnarray}
are expressed in terms of the QCD gluon and 4-quark condansates. There is no
consensus on the value of the 4-quark condensate, in particular the vacuum
saturation approximation $\langle (\bar{q}q)^{2}\rangle =\langle \bar{q}%
q\rangle ^{2}$ is likely to be badly violated \cite{9}. Fortunately the
contribution of the non-perturbative condensates to our sum rules is quite
small which renders our results insensitive to their exact values.

As stated in the introduction, our purpose is to extrapolate expression (\ref
{2.2}) to $t=0$ and compare with the experiemtal measurment of $F(0,Q^{2})$
done by CELLO \cite{2} and CLIO \cite{3}.

At low values of $t$, $F(t,Q^{2})$ is dominated by the $\rho -$meson pole 
\begin{equation}
F(t,Q^{2})=\frac{g(Q^{2})}{(t-m_{\rho }^{2})}+\cdots  \tag{2.5}  \label{2.5}
\end{equation}

Furthermore, $F(t,Q^{2})$ is an analytic function in the complex $t-$plane
with a cut on the positive real axis running from threshold $t=4m_{\pi }^{2}$
to infinity. Consider the integral 
\begin{equation}
F(0,Q^{2})=\frac{1}{2\pi i}\int_{c}dt\,K_{N}(t)F(t,Q^{2})  \tag{2.6}
\label{2.6}
\end{equation}
where $c$ is the contour consisting of a circle of large radius $R$ and two
straight lines parallel to and lying just above and just below the cut. 
\begin{equation}
K_{N}(t)=\frac{1}{t}-\frac{1}{R}\sum_{n=0}^{N}a_{n}\left( \frac{t}{R}\right)
^{n}  \tag{2.7}  \label{2.7}
\end{equation}
and the coefficients $a_{n}$ are up to now arbitrary constants. If $R$ is
large enough, the QCD expression (\ref{2.2}) is expected to hold on the
circle except in the region close to the positive real axis. On the cut and
for low values of $t$, $\rho -$pole dominance, expressed by eq. (\ref{2.5})
adequatly represents the amplitude. We now choose the coefficients $a_{n}$
defining the kernel eq. (\ref{2.7}) such that $K_{N}(t)$ as well as its
derivatives up to order $N$ vanish at $t=R$ in order to annihilate the
integrand in the integration region not covered by vector meson dominance
and where the QCD\ expression cannot be trusted. This is achieved by taking $%
\sum_{n=0}^{N}a_{n}\left( \frac{t}{R}\right) ^{n}$ to coincide with the
Taylor expansion of $\frac{1}{t}$ about $t=R$ truncated at $N$. The integral
(\ref{2.6}) then yields 
\begin{equation}
F(0,Q^{2})=-g(Q^{2})K_{N}(m_{\rho }^{2})+\frac{1}{2\pi i}\oint
dt\,K_{N}(t)F_{p}(t,Q^{2})+c_{0}-\frac{a_{0}c_{2}}{R}-\frac{a_{1}c_{4}}{R^{2}%
}  \tag{2.8}  \label{2.8}
\end{equation}

The duality radius in the $\rho -$meson channel $R$ is taken to be $R\simeq
1.5\mathrm{GeV}^{2}$ \cite{5}. In order to determine $F(0,Q^{2})$ we take $%
N=1$ and $N=2$ successively in eq. (\ref{2.8}) and perform the integral over
the circle of radius $R$ in the complex $t-$plane. The result is shown in
fig.1 and agreement with the data is seen to be quite good.

The deep inelastic limit $Q^{2}\gg R$ is readily obtained, in this limit the
pole of $F_{p}(t,Q^{2})$ lies outside the integration contour and only the $%
1/t$ term in $K_{N}(t)$ contributes so that 
\begin{equation}
F(0,Q^{2})\underset{Q^{2}\rightarrow \infty }{\rightarrow }%
-g(Q^{2})K_{N}(m_{\rho }^{2})+M^{2}Q^{2}\int_{0}^{R_{0}/M^{2}}dy\frac{\exp
(-y)}{(R+Q^{2}+M^{2}y)^{2}}  \tag{2.9}  \label{2.9}
\end{equation}

The fact that the r.h.s. above should not depend on $N$ implies that $%
Q^{2}g(Q^{2})_{\overrightarrow{Q^{2}\rightarrow \infty }}0$ as expected and
that 
\begin{equation}
\frac{Q^{2}F(0,Q^{2})}{8\pi ^{2}f_{\pi }^{2}}\underset{Q^{2}\rightarrow
\infty }{\rightarrow }\frac{M^{2}\left( 1-\exp \left( -\frac{R_{0}}{M^{2}}%
\right) \right) }{8\pi ^{2}f_{\pi }^{2}}\simeq .75  \tag{2.10}  \label{2.10}
\end{equation}

The above result can be compared to values obtained from eq. (\ref{1.1}) if
one uses asymptotic forms of the pion distribution amplitude as given by
Brodsky and Lepage \cite{10} and by Chernyak and Zhitnitsky \cite{11} 
\begin{eqnarray}
&&\frac{Q^{2}F^{BL}(0,Q^{2})}{8\pi ^{2}f_{\pi }^{2}}\underset{%
Q^{2}\rightarrow \infty }{\rightarrow }1  \notag \\
&&\frac{Q^{2}F^{CZ}(0,Q^{2})}{8\pi ^{2}f_{\pi }^{2}}\underset{%
Q^{2}\rightarrow \infty }{\rightarrow }\frac{5}{3}  \TCItag{2.11}
\label{2.11}
\end{eqnarray}

The limiting value given by eq. (\ref{2.10}), together with the bound at $%
Q^{2}=0$ imposed by the anomaly relation $F_{\gamma ^{\ast }\gamma ^{\ast
}\pi ^{0}}(0,0)=\frac{1}{\sqrt{2}\pi f_{\pi }}$ can be accomodated by the
interpolation formula 
\begin{equation}
\frac{Q^{2}F_{\gamma ^{\ast }\gamma ^{\ast }\pi ^{0}}(Q^{2})}{8\pi
^{2}f_{\pi }^{2}}=\frac{Q^{2}}{Q^{2}+\mu ^{2}}\cdot \frac{1}{8\pi ^{2}f_{\pi
}^{2}}  \tag{2.12}  \label{2.12}
\end{equation}
with $\mu ^{2}=M^{2}\left( 1-\exp \left( -\frac{R_{0}}{M^{2}}\right) \right)
=.5\mathrm{GeV}^{2}$

Expression (\ref{2.12}) yields values of the form factor very close to the
ones obtained analytically from eq. \ref{2.8}.

\section{The Anomaly Relation and $\protect\omega -\protect\rho -\protect\pi 
$ coupling}

The anomaly relation \cite{6} determines the form factor when both photons
are on their mass-shell 
\begin{equation}
F(0,0)=\sqrt{2}\pi f_{\pi }F_{\gamma \gamma \pi ^{0}}(0,0)=1  \tag{3.1}
\label{3.1}
\end{equation}

For the symmetric case $q_{1}^{2}=q_{2}^{2}=t$, the asymptotic expression of
Radyushkin and Ruskov \cite{4} takes the form 
\begin{equation}
F^{QCD}(t)=-2\int_{0}^{R_{0}}ds\exp \left( -\frac{s}{M^{2}}\right)
\int_{0}^{1}dx\frac{x\bar{x}t^{2}}{(t-x\bar{x}s)^{3}}+\frac{c_{2}}{t}+\frac{%
c_{4}}{t^{2}}+\frac{c_{6}}{t^{3}}+\cdots  \tag{3.2}  \label{3.2}
\end{equation}
with 
\begin{eqnarray}
c_{2}(M^{2}) &=&-\left[ \frac{\pi ^{2}}{9M^{2}}\langle \frac{\alpha _{S}GG}{%
\pi }\rangle +\frac{704}{203}\pi ^{2}\alpha _{S}\langle (\bar{q}%
q)^{2}\rangle \right]  \notag \\
c_{4}(M^{2}) &=&-\frac{\pi ^{2}}{9}\langle \frac{\alpha _{S}GG}{\pi }\rangle
\notag \\
c_{6}(M^{2}) &=&-\frac{128}{27}\pi ^{2}\alpha _{S}\langle (\bar{q}%
q)^{2}\rangle  \TCItag{3.3}  \label{3.3}
\end{eqnarray}

As before we start from the integral 
\begin{equation*}
F(0)=\frac{1}{2\pi i}\oint dtK_{N}(t)F(t)
\end{equation*}
with $K_{N}(t)$ defined by eq. \ref{2.7}. In this case, the amplitude $F(t)$
possesses a double as well as a single pole at $t=m_{\rho }^{2}$ because the 
$\rho $ and $\omega $ mesons are degenerate in mass, i.e. in the low energy
region 
\begin{equation}
F(t)=\frac{f}{\left( t-m_{\rho }^{2}\right) ^{2}}+\frac{g}{\left( t-m_{\rho
}^{2}\right) }+\cdots  \tag{3.4}  \label{3.4}
\end{equation}

One obtains then 
\begin{equation}
F(0)=-fK_{N}^{\prime }(m_{\rho }^{2})-gK_{N}(m_{\rho }^{2})+\frac{1}{2\pi i}%
\oint dtK_{N}(t)F_{p}(t)-\frac{a_{0}c_{2}}{R}-\frac{a_{1}c_{4}}{R^{2}}-\frac{%
a_{2}c_{6}}{R^{3}}  \tag{3.5}  \label{3.5}
\end{equation}
with the perturbative part $F_{p}(t)$ given by the first term in the r.h.s.
of eq. (\ref{3.2}).

The residue $g$ which describes the transition $\pi +$vector meson$%
\rightarrow $continuum is unknown. The residue $f$ is related to $G_{\omega
\rho \pi }$, the $\omega \rho \pi $ coupling constant, 
\begin{equation}
f=8\pi ^{2}f_{\pi }\frac{m_{\rho }^{2}}{g_{\rho }}\frac{m_{\omega }^{2}}{%
g_{\omega }}G_{\omega \rho \pi }  \tag{3.6}  \label{3.6}
\end{equation}
where $\frac{m_{v}^{2}}{g_{v}}$ denote the coupling of the vector currents
to the corresponding vector mesons.

Taking $N=0$ and $N=1$ in eq. \ref{3.5} with $F(0)=1$ eliminates $g$ and
yields\qquad 
\begin{equation}
G_{\omega \rho \pi }=12.2\mathrm{GeV}^{-1}  \tag{3.7}  \label{3.7}
\end{equation}
a value close to the one given by vector meson dominance $G_{\omega \rho \pi
}=11\mathrm{GeV}^{-1}$. A check of the stability of the method is provided
by the taking $N=2$ with the values obtained for $f$ and $g$ inserted in the
r.h.s. of eq. \ref{3.5} to evaluate $F(0)$, this yields\footnote{%
This corresponds to the factorisation value $\langle (\bar{q}q)^{2}\rangle
=\langle \bar{q}q\rangle ^{2}$. The case $\langle (\bar{q}q)^{2}\rangle
=3\langle \bar{q}q\rangle ^{2}$, e.g., yields $F(0)=.98$.} 
\begin{equation}
F(0)=.96  \tag{3.8}  \label{3.8}
\end{equation}

\section{Pion charge and pion charge radius}

\-In this paragraph the effectiveness and stability of the method used above
is tested in the calculation of the pion charge and charge radius. The
electromagnetic form factor of the pion is 
\begin{equation}
\langle \pi ^{+}(p_{2})\mid V_{\mu }^{em}\mid \pi ^{+}(p_{1})\rangle
=F(t)(p_{1}+p_{2})_{\mu }  \tag{4.1}  \label{4.1}
\end{equation}
and 
\begin{equation}
F(0)=1  \tag{4.2}  \label{4.2}
\end{equation}
is the charge of the pion.

The asymptotic behaviour of\ $F(t)$\ was obtained some time ago by Ioffe and
Smilga \cite{7} and by Nesterenko and Radyushkin \cite{8} 
\begin{equation}
F^{QCD}(t)=-16\pi f_{\pi }^{2}\frac{\alpha _{S}(t)}{t}+F_{p}(t)+\frac{c_{4}}{%
M^{2}}+\frac{c_{6}}{M^{4}}+\frac{2}{13}\frac{c_{6}}{M^{6}}t+\cdots  \tag{4.3}
\label{4.3}
\end{equation}
the first term is asymptotically leading \cite{1} and \ 
\begin{equation}
F_{p}(t)=\frac{1}{4\pi ^{2}f_{\pi }^{2}}\left[ 3\int_{0}^{R_{0}}\frac{\exp
\left( -\frac{x}{M^{2}}\right) x^{2}dx}{(t-2x)^{2}}+4\int_{0}^{R_{0}}\frac{%
\exp \left( -\frac{x}{M^{2}}\right) x^{3}dx}{(t-2x)^{3}}\right]  \tag{4.4}
\label{4.4}
\end{equation}
\begin{equation*}
c_{4}=\frac{1}{24f_{\pi }^{2}}\langle \frac{\alpha _{S}G^{2}}{\pi }\rangle
\;,\;c_{6}=\frac{104\pi }{81f_{\pi }^{2}}\langle (\bar{q}q)^{2}\rangle
\end{equation*}
$R_{0}\simeq 1.2\mathrm{GeV}^{2},\;M^{2}\simeq 1\mathrm{GeV}^{2}$ have been
obtained from stability considerations.

As before consider 
\begin{eqnarray}
F(0) &=&\frac{1}{2\pi i}\int_{c}dtK_{N}(t)F(t)  \notag \\
K_{N}(t) &=&\frac{1}{t}-\frac{1}{R}\sum a_{n}\left( \frac{t}{R}\right)
^{n}\;\;\;\text{or}  \notag \\
F(0) &=&f_{\rho }G_{\rho \pi \pi }K_{N}(m_{\rho }^{2})+\frac{1}{2\pi i}\oint
dtK_{N}(t)F^{QCD}(t)  \TCItag{4.5}  \label{4.5}
\end{eqnarray}
Using standard values for the condensates yields 
\begin{eqnarray}
F(0) &=&.95\;\;\;\;\;\;\;N=0  \notag \\
F(0) &=&1.15\;\;\;\;\;\;N=1  \notag \\
F(0) &=&1.15\;\;\;\;\;\;N=2  \TCItag{4.6}  \label{4.6}
\end{eqnarray}
in reasonably good agreement with experiment.

\ \ The derivative $F^{\prime }(0)$ yields the charge radius 
\begin{equation}
F^{\prime }(0)=\frac{1}{2\pi i}\int_{c}dtK_{N}(t)F(t)  \tag{4.7}  \label{4.7}
\end{equation}
where now 
\begin{equation}
K_{N}(t)=\frac{1}{t^{2}}-\frac{1}{R^{2}}\sum^{N}a_{n}\left( \frac{t}{R}%
\right) ^{n}  \tag{4.8}  \label{4.8}
\end{equation}
and once more the $a_{n}$'s are chosen so as to annihilate $K_{N}$ and it s
derivatives up to order $N$\ at $t=R$. 
\begin{equation}
F^{\prime }(0)=f_{\rho }G_{\rho \pi \pi }K_{N}(m_{\rho }^{2})+\frac{1}{2\pi i%
}\oint dtK_{N}(t)F^{QCD}(t)  \tag{4.9}  \label{4.9}
\end{equation}
the result is 
\begin{eqnarray}
F^{\prime }(0) &=&1.7\mathrm{GeV}^{2}\;\;\;\;N=0  \notag \\
F^{\prime }(0) &=&1.93\mathrm{GeV}^{2}\;\;\;N=1  \notag \\
F^{\prime }(0) &=&2.05\mathrm{GeV}^{2}\;\;\;N=2
\end{eqnarray}
corresponding to a charge radius 
\begin{equation}
\langle r_{\pi }^{2}\rangle =\left( .47\pm .04\right) fm^{2}  \tag{4.11}
\label{4.11}
\end{equation}
which compares quite well with the experimental value $\langle r_{\pi
}^{2}\rangle =.44fm^{2}$. \ \ \ \ \ \ \ \ \ \ \ \ \ \ \ \ \ \ \ \ \ \ \ \ \ .

\ \ In the examples treated above the contribution of the condensates is
quite small and for the chosen values of $N$ \ unknown higher condensates do
not contribute.

\bigskip 

I thank the Academie de Versailles for its generous support and the IPN,
Orsay for their kind hospitality.

\bigskip 

\pagebreak

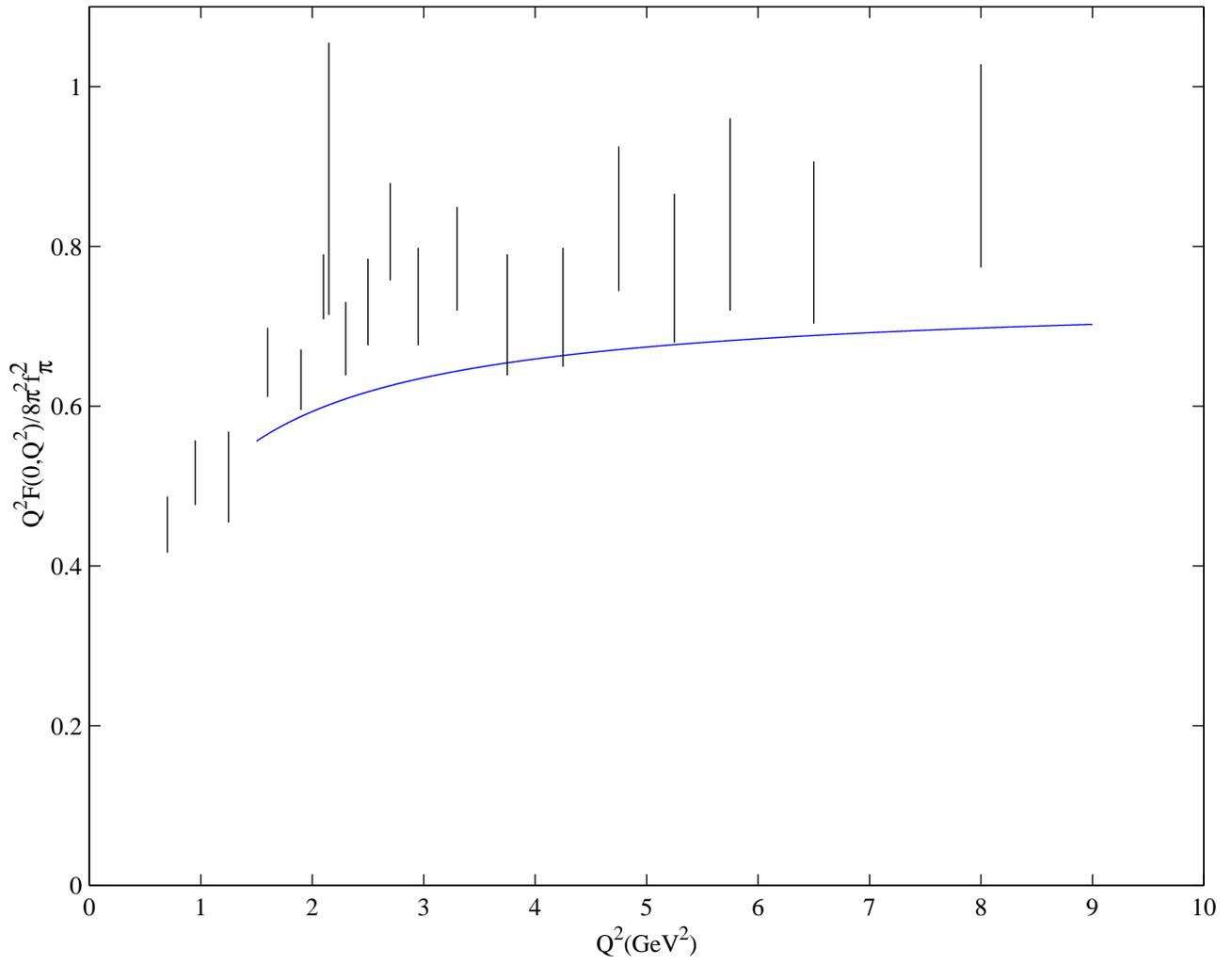
\end{document}